\documentclass[prl,showpacs,twocolumn]{revtex4}
\usepackage{graphicx}
\begin{document}
\title{Observation of molecules produced from a Bose-Einstein condensate}
\author{Stephan D{\"u}rr, Thomas Volz, Andreas Marte, and Gerhard Rempe}
\affiliation{Max-Planck-Institut f{\"u}r Quantenoptik, Hans-Kopfermann-Str.\ 1, 85748 Garching, Germany}
%
\hyphenation{Fesh-bach}
\begin{abstract}
Molecules are created from a Bose-Einstein condensate of atomic $^{87}$Rb using a Feshbach resonance. A Stern-Gerlach field is applied, in order to spatially separate the molecules from the remaining atoms. For detection, the molecules are converted back into atoms, again using the Feshbach resonance. The measured position of the molecules yields their magnetic moment. This quantity strongly depends on the magnetic field, thus revealing an avoided crossing of two bound states at a field value slightly below the Feshbach resonance. This avoided crossing is exploited to trap the molecules in one dimension.
\end{abstract}
\pacs{03.75.Nt, 34.50.-s, 33.15.Kr}
%
\maketitle

In recent years, Bose-Einstein condensation (BEC) has been demonstrated for a variety of atomic species. Creating a BEC of a molecular gas, however, is still an open goal. While cooling a molecular gas is difficult \cite{weinstein:98}, the creation of a molecular BEC from an atomic BEC seems promising. First experiments in this direction have shown the coupling of BEC atoms to a molecular state using either photo-association
\cite{wynar:00,gerton:00,mckenzie:02} or a Feshbach resonance \cite{donley:02}. But while these experiments observed atoms disappear, they did not directly detect the molecular fraction. Very recently, the production of cold molecules from a Fermi gas of atoms was reported \cite{regal:03}, and there, the fraction of molecules could be detected using a spectroscopic technique. Here we report the production of molecules from an atomic BEC using a Feshbach resonance. An alternative observation technique for the molecules is introduced, namely a Stern-Gerlach experiment in which molecules and atoms are spatially separated from each other.

\begin{figure} [b]
\includegraphics[width=.3\textwidth]{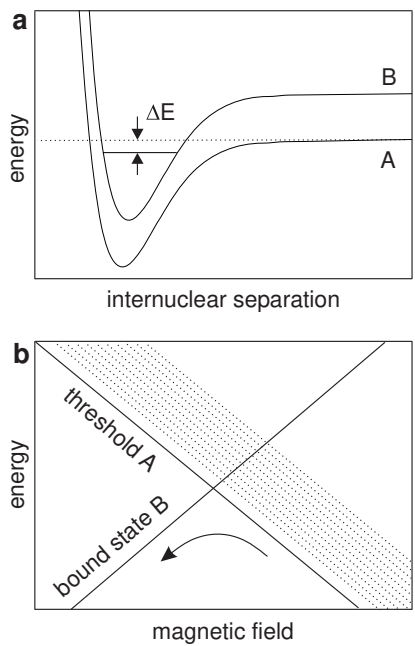}
\caption{\label{fig-feshbach} 
Theory of Feshbach resonances.
(a) 
Molecular potentials involved in a Feshbach resonance. Potential A corresponds to the entrance channel and potential B to a closed channel of the scattering process. Potential B supports a bound state (solid straight line), to which the incoming wave function can couple. The energy of an incoming pair of cold atoms is very close to the dissociation threshold (dotted line). For $\Delta E \rightarrow 0$, the population of the bound state is resonantly enhanced.
(b)
Creation of cold molecules using a Feshbach resonance. The energy of the dissociation threshold and the molecular bound state are shown as a function of magnetic field (solid lines). An atomic BEC can be converted into molecules using an adiabatic rapid passage. The experiment uses a negative ramp speed ($dB/dt<0$), as indicated by the arrow. For positive ramp speed, the existence of non-condensate states of atom pairs (dotted lines) would prevent the build-up of a large molecular fraction. 
 }
\end{figure}

A Feshbach resonance arises from the resonant coupling of an entrance channel to a bound state during a collision. The corresponding potential curves are illustrated in Fig.~\ref{fig-feshbach}a. For simplicity, only the entrance channel A and one closed channel B are shown. Channel B supports numerous bound states of which only one is indicated. If an appropriate interaction couples the two channels, the collision partners can populate the bound state at small internuclear distance. This population is resonantly enhanced, if the bound-state energy approaches the dissociation threshold. This is called a Feshbach resonance \cite{stwalley:76,tiesinga:93}. In order to tune the system into resonance, the energies of the channels A and B must be shifted with respect to each other. Since their magnetic moments are generally different, this can be achieved by applying a magnetic field. In the vicinity of a Feshbach resonance, elastic and inelastic scattering properties change drastically. Feshbach resonances have been observed experimentally in various alkali atoms \cite{inouye:98,courteille:98,roberts:98,vuletic:99,loftus:02,khayokovich:02,strecker:02,marte:02,dieckmann:02,jochim:02,ohara:02,weber:03}.

A Feshbach resonance offers the intriguing possibility to create cold molecules by using an adiabatic rapid passage. This technique is illustrated in Fig.~\ref{fig-feshbach}b, where the energy of the dissociation threshold of the entrance channel  and the bound-state energy are shown as a function of an applied magnetic field $B$. The Feshbach resonance occurs at the intersection and can be described in terms of an avoided crossing. When ramping slowly through the avoided crossing, pairs of atoms can be converted adiabatically into molecules. Starting from an atomic BEC, one might even create a molecular BEC. But, the direction of the ramp is crucial. For the situation shown in Fig.~\ref{fig-feshbach}b, a negative ramp speed ($dB/dt<0$) is predicted to create a significant population of a molecular BEC, but this is not the case for a positive ramp speed \cite{abeelen:99,timmermans:99,mies:00,yurovsky:03,mackie:physics0210131}. Therefore, a negative ramp speed is chosen in the experiment.

The magnitude of the ramp speed is, of course, a crucial parameter in an adiabatic rapid passage. With a fast ramp, the conversion is no longer adiabatic, thus creating only few molecules. With a slow ramp, however, the finite lifetime of the states involved becomes an issue. In the system under investigation here, the molecules are in a highly excited vibrational state which is probably short-lived, because a collision with an atom or another molecule is likely to be inelastic. Therefore a complete conversion of an atomic BEC into molecules is generally difficult and not achieved in this experiment. Such a ramp has also been applied to a degenerate gas of fermionic atoms \cite{regal:03,cubizolles:cond-mat0308018} and to a BEC of $^{85}$Rb \cite{hodby:DAMOP}.

The experiment described here uses bosonic atoms and begins with the preparation of a BEC of $^{87}$Rb atoms that is then transferred into a crossed dipole trap, as described in detail in Refs.~\cite{marte:02,volz:03}. Typical atom numbers each in the BEC and in the surrounding thermal cloud are $10^5$, leading to a peak density of $2 \times 10^{14}$~cm$^{-3}$ at trap frequencies of $2 \pi \times (50,120,170)$~ Hz. The atoms populate the spin state $|f,m_f\rangle = |1,1\rangle$, in which a Feshbach resonance occurs at 1007.40~G \cite{volz:03}. Next, a homogeneous magnetic field of $\sim 1008$~G is applied. During the turn on of this field, the system crosses the Feshbach resonance at such a large ramp speed ($\sim 40$~G/ms) that the resonance has little effect on the atoms.

Next, the dipole trap is switched off and the rest of the experimental sequence takes place with the particles in free flight. The magnetic field is now ramped more slowly (typically $-1$~G/ms) through the Feshbach resonance in order to create molecules. The field is held at a certain value $B_{hold}$ for 3~ms. During this time, an additional Stern-Gerlach field is applied with a typical magnetic-field gradient of 23~G/cm. This extra field accelerates the atoms and molecules differently because of the difference in their magnetic moments. The Stern-Gerlach field is then turned off and the magnetic field is ramped back to $\sim 1008$~G at a slow ramp speed of typically 1~G/ms. This converts some (or all) molecules back to atoms. The magnetic field is then turned off completely, again crossing the resonance so rapidly that this has little effect. The atoms fly freely for another 3~ms, to allow the two clouds to separate further in space. Finally, an absorption image is taken with laser light that is near-resonant with an atomic transition. Atoms that were converted into molecules but never converted back cannot be seen with this technique.

\begin{figure} [b]
\includegraphics[width=.25\textwidth]{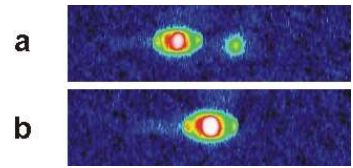}
\caption{ \label{fig-two-clouds}
(Color online) 
(a)
Image of two atomic clouds. The right cloud was temporarily converted into molecules. A Stern-Gerlach field was applied to separate the molecules (right) from the remaining atoms (left). The molecules were converted back into atoms for imaging. The size of the image is $1.3 \times 0.34$~mm.
(b)
Reference image without application of the Stern-Gerlach field.
 }
\end{figure}

Experimental results for $B_{hold} = 1005.2$~G are shown in Fig.~\ref{fig-two-clouds}a, where two clearly separated atom clouds are visible. With the Stern-Gerlach field applied, the magnetic field decreases from left to right in this figure. For reference, the atomic distribution obtained without the Stern-Gerlach field is shown in Fig.~\ref{fig-two-clouds}b. Therefore, the cloud on the left/right in Fig.~\ref{fig-two-clouds}a must have been in a high/low-field seeking state during application of the Stern-Gerlach field. Hence, the right cloud shows atoms that were converted into low-field seeking molecules during the first crossing of the Feshbach resonance and converted back when ramping back. The left cloud, however, consisted of high-field seeking atoms throughout the whole experiment.

With decreasing speed of the molecule-creation ramp, the number of molecules first increases and then saturates for ramp speeds slower than $\sim 2$~G/ms. With such slow ramps, the right cloud contains $\sim 7$~\% of the initial atom number, the left cloud contains $\sim 30$~\%. The rest of the atoms is missing. This atom loss is $\sim 3$ orders of magnitude faster than expected from the previously measured three-body loss coefficient at the resonance \cite{marte:02}. The missing fraction of $\sim 63$~\% might have been lost due to inelastic collisions (presumably after conversion into molecules) or might have been converted into molecules and not converted back to atoms, thus remaining invisible during detection. The latter seems unlikely, because when varying the ramp that converts molecules back to atoms, the back-converted atom number shows no significant dependence on the ramp speed in the range between 0.3 and 5~G/ms.

If the conversion to molecules is performed before release from the dipole trap, the observed number of molecules is reduced by a factor of $\sim 1.8$. This might be due to faster inelastic collision rates at the high densities in the trap. In the experiment, the cloud is therefore allowed to expand after turning off the dipole trap. After an expansion time of typically 2 to 7~ms, the molecules are produced. This expansion is estimated to reduce the atomic density by a factor between 5 and 70 as compared to the in-trap situation. The observed number of molecules shows no strong density dependence over this range. At first glance, this seems to rule out inelastic collisions as the dominant factor that limits the molecule creation efficiency. However, at reduced density a slower ramp is needed to associate molecules efficiently. This in turn requires more time to be spent during the ramp, therefore allowing for more loss due to inelastic collisions. It is thus not clear, whether a reduced density is expected to achieve an overall improvement in the observed number of molecules. We therefore think that inelastic collisions are the most likely candidate for explaining the small conversion efficiency.

The displacement of the clouds induced by application of the Stern-Gerlach field can be used to extract the magnetic moment of the particles. This displacement is proportional to the force $\vec{F} = -(dE/dB) \times \nabla |\vec{B}|$ exerted by the Stern-Gerlach field. Here, $E$ is the energy of the atomic and molecular state, respectively (see Fig.~\ref{fig-feshbach}b). Note that $dE/dB$ is related to the magnetic moment of the particles. For the atoms, the extracted experimental value of $dE/dB$ is independent of $B_{hold}$ and agrees well with the expectation from the Breit-Rabi formula.

\begin{figure} [b]
\includegraphics[width=.35\textwidth]{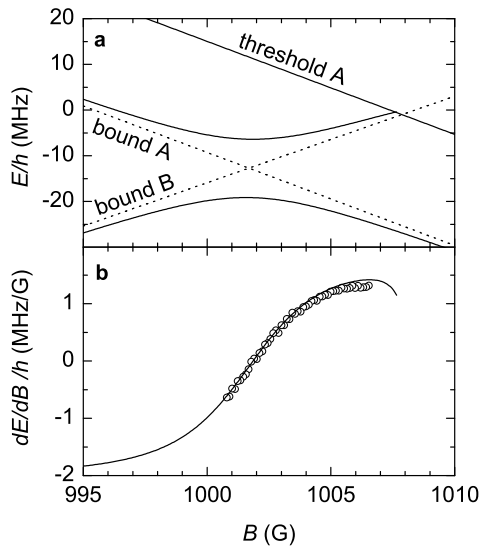}
\caption{ \label{fig-dEdB}
Avoided crossing.
(a)
Energy as a function of magnetic field. The binding energy of the highest-lying bound state in potential A with respect to the dissociation threshold in potential A is $h\times 24$~MHz, independent of the magnetic field. The bound state supported by potential B crosses both these energies. The highest-lying bound state is crossed at 1001.7~G, leading to an avoided crossing. The dissociation threshold is crossed at 1007.4~G creating the Feshbach resonance. The energy splitting in the avoided crossing at 1001.7~G is 13~MHz and therefore much larger than the splitting of the Feshbach resonance crossing, which is not visible on this scale.
(b)
Derivative $dE/dB$ of the upper branch of the avoided crossing. Experimental data for the molecules (circles) extracted from images, such as Fig.~\ref{fig-two-clouds}a, are compared with the theory (solid line) that contains no free fit parameters.
 }
\end{figure}

\begin{figure} [b!]
\includegraphics[width=.27\textwidth]{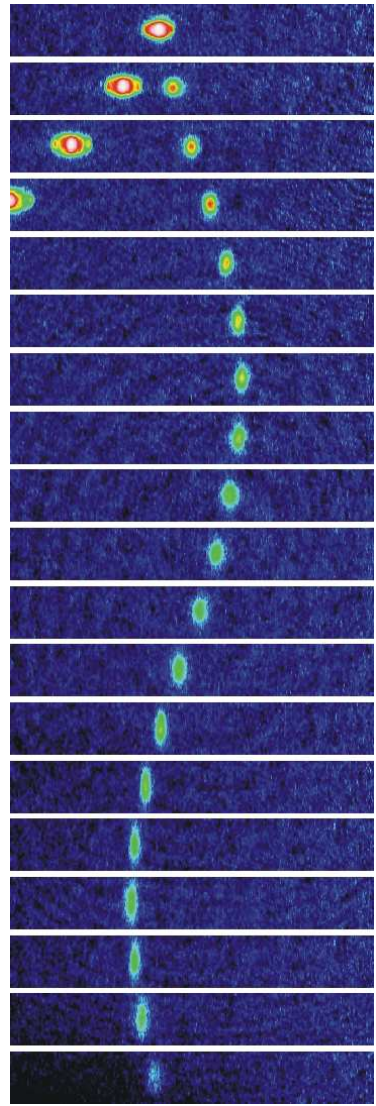}
\caption{ \label{fig-oscillation}
(Color online) Oscillation of molecules. According to Fig.~\ref{fig-dEdB}, the molecules are 1001.7-G seekers. With a magnetic field gradient applied in the horizontal direction, the molecules oscillate around a point in space where the magnetic field equals 1001.7~G. The images were recorded for a series of different durations of the Stern-Gerlach field, ranging from 0 to 18~ms (top to bottom). The observed oscillation frequency of 56~Hz agrees well with theory. The anisotropic expansion of the molecular cloud is due to the fact that the one-dimensional trapping potential prevents the cloud from expanding in the horizontal direction. The atomic cloud (top left) is simply accelerated on a parabola. The size of each image is $1.7 \times 0.24$~mm.
 }
\end{figure}

For the molecules, however, Fig.~\ref{fig-dEdB}b displays a pronounced dependence of $dE/dB$ on $B_{hold}$. This magnetic-field dependence is due to the presence of the highest-lying bound state in entrance channel A. Theoretical results \cite{vanKempen:pers} for the energy of this state are shown in Fig.~\ref{fig-dEdB}a. At a field value of 1001.7~G, this bound state crosses the bound state in potential B that becomes populated when ramping across the Feshbach resonance. Due to the exchange interaction, this is an avoided crossing, in which the molecules adiabatically follow the upper branch. The derivative $dE/dB$ of this upper branch is shown in Fig.~\ref{fig-dEdB}b. The good agreement between theory and experiment clearly proves that molecules are created in the expected ro-vibrational level.

When a molecule passes through the upper branch of the avoided crossing, its vibrational quantum number changes from -5 to -1 (counting from threshold). Correspondingly, the size of the molecule, {\it i.e.}\ the outer turning point of the vibrational state of the nuclei, changes by a factor of $\sim 3$.

The reversal of the sign of $dE/dB$ visible in Fig.~\ref{fig-dEdB}b makes the molecules 1001.7-G seekers, instead of the usual high- or low-field seekers. This can be exploited to trap the molecules by applying an inhomogeneous magnetic field. In the presence of a magnetic-field gradient, the quadratic dependence of $E(B)$ in Fig.~\ref{fig-dEdB}a creates a harmonic confinement for the molecules. The resulting harmonic oscillation is shown in Fig.~\ref{fig-oscillation} for a gradient of 100~G/cm with the molecules initially prepared at 1003.2~G. This corresponds to a one-dimensional trap for molecules. In principle, such a trap could work in three dimensions. Note that this is very different from a standard magnetic trap, where usually low-field seeking particles are trapped at a minimum of $|\vec{B}|$. Here, the molecules are trapped at a specific value of $|\vec{B}|$, which is not a minimum of the field configuration.

In conclusion, the Stern-Gerlach technique used here proves that molecules are created in the expected ro-vibrational level. The measured magnetic moment of the molecules reveals an avoided crossing and agrees well with theory. The avoided crossing is used to trap the molecules in one dimension. Future experiments will aim on the creation of a larger fraction of molecules, investigate heating mechanisms that occur during the magnetic field ramps, and develop tools to investigate the coherence properties of the molecular cloud.

Similar experiments are currently being carried out with a BEC of Cs \cite{herbig:03}. After submission of this manuscript we learned that related experiments were performed with Na \cite{xu:cond-mat0310027} and fermionic $^6$Li \cite{jochim:cond-mat0308095,strecker:03}.

We thank S. Ernst for assistance with the experiment, and E. van Kempen and B. Verhaar for providing numerical results. This work was supported by the European-Union Network ``Cold Quantum Gases".


\end{document}